\documentstyle[12pt]{article}
\newcommand{\EPJ}{Eur. Phys. J. }

\newcommand{\JHEP}{J. High Energy Phys. }

\newcommand{\MPL}{Mod. Phys. Lett. }
\newcommand{\NP}{Nucl. Phys. }
\newcommand{\NJP}{New J. Phys. }
\newcommand{\PR}{Phys. Rev. }
\newcommand{\PRL}{Phys. Rev. Lett. }
\newcommand{\PL}{Phys. Lett. }
\newcommand{\RMP}{Rev. Mod. Phys. }
\newcommand{\ZP}{Z. Phys. }

\newcommand{\VC}{V^{\rm C}}
\newcommand{\VCs}{V^{{\rm C}*}}
\newcommand{\VCd}{V^{{\rm C}\dagger}}
\newcommand{\VCT}{V^{{\rm C}T}}
\newcommand{\VN}{V^{\rm N}}

\begin{document}

\baselineskip=16pt
\pagenumbering{arabic}

\vspace{1.0cm}
\begin{flushright}
NKHEP 07/2006\\
hep-ph/0604016
\end{flushright}

% This version 2 added the last two refs to item [17]

\begin{center}
{\Large\sf Active-sterile neutrino mixing in the absence of bare
active neutrino mass}

\vspace{0.5cm}

{Yi Liao\footnote{liaoy@nankai.edu.cn}}

{\small Department of Physics, Nankai University, Tianjin 300071,
China}

\vspace{1.0cm}

{\bf Abstract}

\end{center}

We investigate a minimal extension of the standard model in which
the only new ingredient is the sterile neutrinos. We do not
introduce extra Higgs multiplets or high dimensional effective
operators to induce mass terms for the active neutrinos, and the
model is renormalizable in itself. We show for arbitrary numbers
of generations and sterile neutrinos that the independent physical
parameters in the leptonic sector are much less than previously
anticipated. For instance, with three active and two sterile
neutrinos, there are four mixing angles and three CP phases in
addition to four non-vanishing neutrino masses. We study
phenomenological implications for tritium beta decay, neutrinoless
double beta decay and neutrino oscillations. For the most natural
see-saw parameters, we find that it is difficult to accommodate in
the model the best-fit values of masses and mixing parameters from
oscillation data no matter whether we include or not the null
short-baseline experiments together with the LSND result. This
implies that if the LSND result is confirmed by MiniBooNE, the
see-saw parameter region of the model with two sterile neutrinos
could be largely excluded.

\begin{flushleft}
PACS: 14.60.Pq, 14.60.St, 23.40.-s
% corresponding to: (PACS 2003 version)
% neutrino mass and mixing; non-standard-model neutrinos, right-handed neutrinos,
% etc; beta decay, double beta decay, electron and muon capture

Keywords: sterile neutrino, lepton mixing, neutrino oscillation,
beta decay

\end{flushleft}

\newpage
\section{Introduction}

The experiments on neutrinos have offered the first piece of
evidence that points to physics beyond the standard model (SM) of
electroweak interactions \cite{review}. The deficits in the solar
electron neutrino flux \cite{solar} and in the ratio of the
atmospheric muon to electron neutrino fluxes \cite{atm} can be
best and most naturally interpreted in terms of neutrino
oscillations \cite{nuoscill}. These observations have been
confirmed by experiments at accelerators and reactors
\cite{k2k,kamland}. They imply that the three neutrinos are
non-degenerate and interact non-diagonally with the charged
leptons. Nevertheless, the controversial circumstance in the short
baseline (SBL) experiments seems to indicate that the complete
picture of the neutrino sector may be richer than with the three
ordinary neutrinos: while the LSND experiment claimed to observe a
statistically significant signal \cite{lsnd} that cannot be
accommodated in the three neutrino scheme of solar and atmospheric
experiments due to three very different mass gaps, the signal was
not confirmed by other SBL experiments \cite{nsbl,chooz}. This
situation will hopefully be clarified in the near future by the
MiniBooNE experiment \cite{boone}.

The controversy at SBL experiments has stimulated a lot of
theoretical attempts. Some of them appeal to certain drastically
new physics like CPT violation \cite{cpt}, CPT violating quantum
decoherence \cite{decoh}, or extra dimensions \cite{xdim}, to
mention a few. A more conventional approach is to introduce
additional neutrinos to accommodate the observed at least three
mass gaps. Since the number of SM-like neutrinos is severely
constrained by electroweak measurements to be three, these new
neutrinos must be neutral with respect to the SM gauge groups;
i.e., they must be sterile neutrinos. The concept of sterile
neutrinos was introduced earlier in the other context
\cite{sterile}. The most economical scenario with one sterile
neutrino has been extensively studied in the literature
\cite{4nu}. It is generally difficult, if not impossible, to
explain all data in this scenario either because of the rejection
of significant involvement of a sterile neutrino by the solar and
atmospheric data or because of the tension existing between the
positive SBL result at LSND and the negative one at all others.
The next simplest would be to add two sterile neutrinos, the
so-called (3+2) scenario. Sorel {\it et al}. have assessed
carefully the compatibility of all SBL experiments, and found that
the (3+2) scenario fits the SBL data significantly better than the
one with a single sterile neutrino \cite{sorel}. From the model
building point of view, although it is simplest to add one sterile
neutrino, there are viable models that contain two or three
sterile neutrinos \cite{5nu}. It has to be left for experiments to
decide which is actually realized in Nature.

In this paper, we consider a minimal extension of SM that could
potentially accommodate the neutrino data. It is minimal in the
sense that only the neutrino sector is extended by adding some
sterile neutrinos. We do not introduce additional Higgs particles
or higher dimensional effective operators from some high energy
scale. In particular, there are no bare masses for the ordinary
active neutrinos. The model so extended preserves
renormalizability in itself. In an earlier work \cite{liao} we
found that such a model has a strictly constrained leptonic sector
which contains much less physical parameters than previously
anticipated \cite{sv80a,sv80b}. With one sterile neutrino, for
instance, there are only two mixing angles in addition to two
non-vanishing neutrino masses, and the leptonic sector preserves
CP symmetry automatically. Furthermore, the ratio of the two
masses also appears in the mixing matrices, which makes the two
otherwise independent mixing angles less effective in generating
an experimentally favored mixing matrix than they would have been.
The resulting model is even incapable of interpreting the solar
and atmospheric data \cite{liao,aguila}. In this work, we
investigate the parametrization of the leptonic sector for any
numbers of generations and sterile neutrinos. Since the number of
physical parameters is much less than previously expected, it
becomes numerically manipulatable to study its implications on
neutrino experiments even with two or three sterile neutrinos. For
instance, with two sterile neutrinos, there are four mixing angles
and three CP violating phases in addition to four non-vanishing
neutrino masses, and the model could thus become viable to
accommodate the neutrino data including the SBL experiments.

The paper is organized as follows. In the next section, we study
the leptonic sector of the minimally extended model, and count in
particular the number of physical parameters contained in it for
any numbers of generations and sterile neutrinos. To get prepared
for phenomenological analyses, we work out analytically in section
3 the neutrino spectrum and leptonic mixing matrices for the most
natural see-saw region in parameter space \cite{seesaw}. Then, in
section 4 we discuss the phenomenological implications in the case
of two sterile neutrinos. We do not attempt here a sophisticated
statistical assessment; instead, with the fitting result by Sorel
{\it et al}. for the general (3+2) scenario in mind, we consider
whether it is possible to accommodate their result in the above
mentioned parameter region of our minimal model. The answer turns
out to be negative and even worse: even if we ignore the tension
between the positive and negative SBL experiments, it is difficult
to include the LSND result in the see-saw region of the model. We
summarize our results in the last section and mention briefly
further work worthy to do with the model.

\section{Mixing matrices for any numbers of active and sterile neutrinos}
In this section we first describe the leptonic sector of the
$n$-generation SM extended by $n_0$ sterile neutrinos. This is the
minimal framework that can accommodate neutrino mass and mixing
while preserving renormalizability of SM. By standardizing the
neutrino mass matrices, we count independent physical parameters
contained in the leptonic sector. Diagonalization of the leptonic
mass matrices then yields the mixing matrices in the charged and
neutral current interactions. For $n>n_0$, we show that the number
of independent physical parameters can be further reduced due to
the appearance of massless neutrinos.

\subsection{Setup of the model}

The only new fields compared to SM are the $n_0$ sterile neutrinos
that we choose to be right-handed without loss of generality,
$s_{Rx}$, $x=1,\cdots,n_0$. The model contains as usual the $n$
generations of the lepton doublets, $F_{La}=(n_{La},f_{La})^T$,
and of the charged lepton singlets, $f_{Ra}$, $a=1,\cdots,n$. Here
$L,R$ refer to the left- and right-handed projections of the
fields.

Since the sterile neutrinos are neutral under $SU(2)_L\times
U(1)_Y$ by definition, they are allowed to have bare mass terms of
Majorana type,
\begin{equation}
-{\cal L}_{s_R}=\frac{1}{2}M_{xy}\overline{s_{Rx}^C}s_{Ry}
+\frac{1}{2}M^*_{xy}\overline{s_{Ry}}s_{Rx}^C
\end{equation}
where $\psi^C={\cal C}\gamma^0\psi^*$ stands for the
charge-conjugate field of $\psi$ with ${\cal C}=i\gamma^0\gamma^2$
satisfying ${\cal C}=-{\cal C}^{\dagger}=-{\cal C}^T=-{\cal
C}^{-1}$ and ${\cal C}\gamma^{\mu T}{\cal C}=\gamma^{\mu}$. We
denote $s_R^C=(s_R)^C$ for brevity. The $n_0\times n_0$ complex
matrix $M$ is symmetric due to anticommutativity of fermion
fields, but is otherwise general. The presence of sterile
neutrinos also introduces the mixing mass terms between active and
sterile neutrinos through the Yukawa interactions,
\begin{equation}
-{\cal L}_{\rm Y}=y_{ab}^f\overline{F_{La}}\varphi f_{Rb}
+y_{ax}^n\overline{F_{La}}\tilde{\varphi}s_{Rx}+{\rm h.c.}
\end{equation}
where $\varphi$ is the Higgs doublet field that develops a vacuum
expectation value, $\langle\varphi\rangle=(0,1)^Tv/\sqrt{2}$, and
$\tilde{\varphi}=i\sigma^2\varphi^*$. Note that no bare mass terms
are allowed for the active neutrinos in this minimal extension of
SM. The lepton mass terms are summarized by
\begin{equation}
-{\cal L}_{\rm m}= \left[\overline{f_L}m_f f_R+\overline{n_L}Ds_R
+{\rm h.c.}\right] +\frac{1}{2}\left[\overline{s_R^C}Ms_R+{\rm
h.c.}\right]
\end{equation}
where $m_f=y^fv/\sqrt{2}$ and $D=y^nv/\sqrt{2}$ are $n\times n$
and $n\times n_0$ complex matrices respectively.

\subsection{Standardization and diagonalization of mass matrices}

To facilitate counting the independent physical parameters
contained in the leptonic mixing matrices after diagonalizing the
lepton mass matrices, we first convert the neutrino mass matrices
into a standard form. Since $M$ is symmetric, it can be
diagonalized by a unitary transformation $s_R\to Y_0s_R$ to the
real positive $Y_0^TMY_0=M_{\rm diag}={\rm
diag}(r_1,\cdots,r_{n_0})$. The only other change in the total
Lagrangian occurs in ${\cal L}_{\rm m}$: $D\to DY_0$. Denote
$DY_0=(a_1,\cdots,a_{n_0})$ where $a_x$ ($x=1,\cdots,n_0$) are
general $n$-component complex vectors. Making a unitary
transformation $n_L\to X^{\dagger}n_L$, we convert $DY_0$ into
$XDY_0=D_{\triangle}$ which has a zero triangle at its upper-left
corner, and when $n>n_0$ an additional $(n-n_0)\times n_0$ zero
rectangle over the triangle. To keep the partnership of $n_L$ and
$f_L$, we also transform $f\to X^{\dagger}f$ so that the only
other change is again in ${\cal L}_{\rm m}$ which now becomes
\begin{equation}
-{\cal L}_{\rm m}= \left[\overline{f_L}Xm_fX^{\dagger}
f_R+\overline{n_L}D_{\triangle}s_R +{\rm h.c.}\right]
+\frac{1}{2}\left[\overline{s_R^C}M_{\rm diag}s_R+{\rm
h.c.}\right]
\end{equation}

The matrix $D_{\triangle}$ is found successively and in a way
applicable to both cases $n>n_0$ and $n\le n_0$. First, we choose
the unitary matrix $X_1$ so that $X_1a_1=(0,\cdots,0,|a_1|)^T$
with $|a_1|=\sqrt{a_1^{\dagger}a_1}$. Invariance of the inner
product of vectors implies that
$(X_1a_x)_n=\frac{a_1^{\dagger}a_x}{|a_1|}$ which determines all
entries in the $n$-th row of $X_1(a_1,\cdots,a_{n_0})$. The first
$(n-1)$ components of the column vector $X_1a_x$ are normalized to
$\sqrt{a_x^{\dagger}P_1a_x}$ with the projector
$P_1=1-\frac{a_1a_1^{\dagger}}{a_1^{\dagger}a_1}$. Next, we choose
the unitary matrix $X_2$ that leaves all entries in the $n$-th row
of $X_1(a_1,\cdots,a_{n_0})$ untouched and that rotates the first
$(n-1)$ components of $X_1a_2$ to its $(n-1)$-th component.
Invariance of the inner product then determines the $(n-1)$-th row
of $X_2X_1(a_1,\cdots,a_{n_0})$, and so on. The procedure
continues by induction until the $n_{\rm min}$-th column vector
where $n_{\rm min}={\rm min}(n,n_0)$. Defining $a_0=0$ and the
projectors
\begin{equation}
P_0=1,~
P_x=1-\frac{\left[\displaystyle\prod_{y=0}^{x-1}P_y\right]a_xa_x^{\dagger}
\left[\displaystyle\prod_{y=0}^{x-1}P_y\right]}
{a_x^{\dagger}\left[\displaystyle\prod_{y=0}^{x-1}P_y\right]a_x},
\end{equation}
which have the properties
\begin{equation}
P_x^{\dagger}=P_x,~P_xP_x=P_x,~\left[\prod_{y=1}^xP_y\right]a_x=0,
~[P_x,P_y]=0,
\end{equation}
the end result, $D_{\triangle}=X(a_1,a_2,\cdots,a_{n_0})$ with
$X=X_{n_{\rm min}}\cdots X_1$, has the following non-vanishing
entries,
\begin{equation}
(D_{\triangle})_{ax}=\displaystyle\frac{\displaystyle
a^{\dagger}_{n+1-a} \left[\prod_{y=0}^{n-a}P_y\right]a_x}
{\displaystyle\sqrt{a^{\dagger}_{n+1-a}\left[\prod_{y=0}^{n-a}
P_y\right]a_{n+1-a}}}
\end{equation}
where the indices are restricted to $a\in[1,n]{\rm
~and~}x\in[n+1-a,n_0]$ when $n\le n_0$, and to $x\in[1,n_0]{\rm
~and~}a\in[n+1-x,n]$ when $n>n_0$.

Now we diagonalize the lepton mass matrices. The charged part is
done as usual by the bi-unitary transformations
$f_{L,R}=X_{L,R}\ell_{L,R}$ with
$X_L^{\dagger}(Xm_fX^{\dagger})X_R=m_{\ell}$ being real positive.
The matrix $X_L^{\dagger}$ then appears in the charged current
interactions of leptons. To diagonalize the neutral part, we first
rewrite the Lagrangian in terms of the fields ${n_L^C\choose s_R}$
and their charge conjugates ${n_L\choose s_R^C}$. Then the
neutrino mass terms become
\begin{equation}
-{\cal L}_{\rm m}^{\nu}=\displaystyle
\frac{1}{2}\left(\overline{n_L},\overline{s_R^C}\right)m_n
\left(\begin{array}{c}n_L^C\\s_R
\end{array}\right)+
\frac{1}{2}\left(\overline{n_L^C},\overline{s_R}\right)m^{\dagger}_n
\left(\begin{array}{c}n_L\\s_R^C
\end{array}\right)
\end{equation}
where the $(n+n_0)$-dimensional, symmetric mass matrix in the new
basis is
\begin{equation}
m_n=\left(\begin{array}{cc}0_n&D_{\triangle}\\D^T_{\triangle}&M_{\rm
diag}
\end{array}\right)
\end{equation}
with $0_n$ being the zero matrix of $n$ dimensions. Finally, the
matrix $m_n$ is diagonalized by the unitary transformation
\begin{equation}
\left(\begin{array}{c}n_L^C\\s_R
\end{array}\right)=Y\nu_R
\end{equation}
such that
\begin{equation}
Y^Tm_nY=m_{\nu}
\end{equation}
is real and non-negative. The $n\times(n+n_0)$ submatrix, $y$,
composed of the first $n$ rows of $Y$ appears in the charged and
neutral current interactions of leptons. We shall denote the mass
eigenstate fields of the neutral leptons, $\nu$, by the Latin
indices, $j,k=1,2,\cdots,n+n_0$, and those of the charged leptons,
$\ell$, by the Greek indices, $\alpha,\beta=1,\cdots,n$.
Introducing the Majorana neutrino fields $\nu=\nu_R+\nu_R^C$, the
changes in the total Lagrangian are summarized as follows,
\begin{equation}
\begin{array}{rcl}
{\cal L}_{\rm m}
&=&\displaystyle-m_{\ell_{\beta}}\overline{\ell_\beta}\ell_{\beta}
-\frac{1}{2}m_{\nu_j}\overline{\nu_j}\nu_j\\
{\cal L}_{\rm
CC}&=&\displaystyle\frac{g}{\sqrt{2}}\left[\VC_{\beta
j}W^-_{\mu}\overline{\ell_{L\beta}}\gamma^{\mu}\nu_j +\VCs_{\beta
j}W^+_{\mu}\overline{\nu_j}\gamma^{\mu}\ell_{L\beta} \right]\\
{\cal L}_{\rm NC}^{\nu}&=&\displaystyle\frac{g}{4c_W}Z_{\mu}
\overline{\nu_k}\gamma^{\mu}\left[i{\rm ~Im~}\VN_{kj}
-\gamma^5{\rm ~Re~}\VN_{kj}\right]\nu_j
\end{array}
\end{equation}
where upon redefining $X_L^{\dagger}=V$ for brevity
\begin{equation}
\VC=Vy^*,~\VC \VCd=1_n,~\VN=\VCd\VC=y^Ty^*
\end{equation}

\subsection{Counting of physical parameters}

We note first of all that $D_{\triangle}$ has different zero
textures according to $n>n_0$ or $n\le n_0$. This affects the
number of physical parameters in the model. We start with the
easier case of $n\le n_0$. The matrix $D_{\triangle}$ has $n$ real
entries lying at the rows and columns fulfilling $a+x=n+1$ and
$\frac{1}{2}n(2n_0-n-1)$ complex ones to the right of the real
entries. There are thus $(n+n_0)$ real parameters and
$\frac{1}{2}n(2n_0-n-1)$ complex ones in $m_n$, which are traded
for $(n+n_0)$ neutrino masses, $\frac{1}{2}n(2n_0-n-1)$ mixing
angles and $\frac{1}{2}n(2n_0-n-1)$ CP violation phases in $Y$.
Since for general original matrices $M,~D$ there are no further
unitary transformations on $s_R,~n_L$ that leave the structure of
$m_n$ invariant, all those angles and phases must appear in the
submatrix $y$ of $Y$, and thus in $\VN$ and $\VC$. The additional
parameters introduced to $\VC$ on diagonalizing the charged lepton
masses are counted as usual: $\frac{1}{2}n(n-1)$ angles and
$\frac{1}{2}n(n-1)$ phases after absorbing the $n$ phases by the
charged lepton fields. Again, for general original matrices, they
are not expected to combine with or cancel those in $y$. In
summary, there are $n$ massive charged leptons, $(n+n_0)$ massive
neutrinos, $n(n_0-1)$ mixing angles and $n(n_0-1)$ CP phases in
the charged current matrix $\VC$. Out of those, only
$\frac{1}{2}n(2n_0-n-1)$ angles and $\frac{1}{2}n(2n_0-n-1)$
phases appear in the neutral current matrix $\VN$ of the
neutrinos.

When $n>n_0$, in addition to the indicated zero triangle, the
first $(n-n_0)$ rows of $D_{\triangle}$ also vanish, which
correspond to $(n-n_0)$ massless neutrinos after diagonalization.
Arbitrary unitary transformations, $y_0$, amongst the massless
modes are allowed without changing the matrix $m_n$. Thus, $y$ has
the structure
\begin{equation}
y=\left(\begin{array}{cc}y_0&0_{(n-n_0)\times 2n_0}\\
0_{n_0\times(n-n_0)}&\bar{y}
\end{array}\right),~y_0^{-1}=y_0^{\dagger}
\end{equation}
While both $\bar{y}$ and $y_0$ appear in $\VC$, only $\bar{y}$
appears in $\VN$. In this case, there are $n_0$ real and
$\frac{1}{2}n_0(n_0-1)$ complex parameters in $D_{\triangle}$;
together with $M_{\rm diag}$, they are traded for $2n_0$ neutrino
masses, $\frac{1}{2}n_0(n_0-1)$ mixing angles and
$\frac{1}{2}n_0(n_0-1)$ CP phases in $\bar{y}$ and thus $\VN$.

Now we count the physical parameters contained in $\VC$. An $n$
dimensional unitary matrix $V$ may be parameterized as a product
in any arbitrarily specified order of the $n$ diagonal phase
matrices, $e_{\alpha}(u_{\alpha})$ ($\alpha=1,\cdots,n$), and the
$\frac{1}{2}n(n-1)$ complex rotation matrices in the
$(\alpha,\beta)$ plane,
$\omega_{\alpha\beta}(\theta_{\alpha\beta},\varphi_{\alpha\beta})$
($n\ge\beta>\alpha\ge 1$) \cite{sv80a}. Here $e_{\alpha}(z)$ is
obtained by replacing the $\alpha$-th entry in $1_n$ by the phase
$z$, and
\begin{equation}
\omega_{\alpha\beta}(\theta_{\alpha\beta},\varphi_{\alpha\beta})
=e_{\alpha}(e^{i\varphi_{\alpha\beta}})R_{\alpha\beta}(\theta_{\alpha\beta})
e_{\alpha}(e^{-i\varphi_{\alpha\beta}})
\end{equation}
where $R_{\alpha\beta}(\theta_{\alpha\beta})$ is the usual real
rotation matrix through angle $\theta_{\alpha\beta}$ in the
$(\alpha,\beta)$ plane. We choose the order of products in such a
way that it fits our purpose here:
\begin{equation}
V=V_2V_1V_0
\end{equation}
where $V_0$ is the general unitary matrix in the subspace spanned
by the first $(n-n_0)$ axes, $V_2$ the unitary matrix in the
subspace spanned by the last $n_0$ axes, and $V_1$ is the one
mixing the two subspaces. In $\VC$, $V_0$ can be cancelled by
$y_0^*$ in $y^*$, which remains free until now. The $n_0$ diagonal
phases in $V_2$ can be arranged to its very left to get absorbed
by the charged lepton fields. We are thus left in $\VC$ with the
$\frac{1}{2}n_0(n_0-1)$ complex rotations in $V_2$ and
$(n-n_0)n_0$ ones in $V_1$, in addition to the
$\frac{1}{2}n_0(n_0-1)$ complex parameters contained in $\bar{y}$.

However, not all parameters in $V_1$ are physical. To see this, we
write
\begin{equation}
V_1=\prod_{a,z}\omega_{az}(\theta_{az},\varphi_{az})
\end{equation}
with $a\in[1,n-n_0],~z\in[n-n_0+1,n]$. The order of products is
specified as follows: factors with the same $z$ are grouped
together with $z$ increasing from left to right; within each
group, $a$ also increases from left to right. Denoting for brevity
$\omega_{az}(\theta_{az},\varphi_{az})=\omega_{az}$,
$R_{az}(\theta_{az})=R_{az}$, $e_a(e^{i\varphi_{az}})=e_{az}$ and
$e_a(e^{-i\varphi_{az}})=e_{az}^*$, it is clear that
\begin{equation} \omega_{1,z}\cdots\omega_{n-n_0,z}=
\left[\prod_{a=1}^{n-n_0}e_{az}\right]
\left[\prod_{b=1}^{n-n_0}R_{bz}\right]
\left[\prod_{c=1}^{n-n_0}e_{cz}^*\right]
\end{equation}
Then,
\begin{equation}
\begin{array}{rcl}
V_1&=&\displaystyle\left[\prod_{a=1}^{n-n_0}e_{a,n-n_0+1}\right]
\left[\prod_{b=1}^{n-n_0}R_{b,n-n_0+1}\right]
\left[\prod_{c=1}^{n-n_0}e_{c,n-n_0+1}^*e_{c,n-n_0+2}\right]\\
&\times&\displaystyle
\left[\prod_{b=1}^{n-n_0}R_{b,n-n_0+2}\right]
\left[\prod_{c=1}^{n-n_0}e_{c,n-n_0+2}^*\right]\cdots
\left[\prod_{a=1}^{n-n_0}e_{a,n-1}\right]
\left[\prod_{b=1}^{n-n_0}R_{b,n-1}\right]\\
&\times&\displaystyle
\left[\prod_{c=1}^{n-n_0}e_{c,n-1}^*e_{cn}\right]
\left[\prod_{b=1}^{n-n_0}R_{bn}\right]
\left[\prod_{a=1}^{n-n_0}e_{an}^*\right]
\end{array}
\end{equation}
The left phase matrix in the first line can be pushed through
$V_2$ to be absorbed by the charged lepton fields, while the right
phase matrix in the last line passes through $y$ in which
$y_0=1_{n-n_0}$ to get absorbed by the massless neutrino fields.
This leaves with us the $(n-n_0)(n_0-1)$ phase differences and all
of the real rotations in $V_1$ that cannot be removed. To
summarize, there are $n_0(n-1)$ physical mixing angles and
$n(n_0-1)$ physical CP phases in $\VC$.

There is another way to count the physical parameters in $\VC$. A
general $n\times(n+n_0)$ complex matrix $\VC$ has $2n(n+n_0)$ real
parameters. First, $\VC\VCd=1_n$ offers $n^2$ real constraints.
Second, $\VC$ may be multiplied from the left by an arbitrary
diagonal phase matrix by redefining the charged lepton fields
without affecting physics. This amounts to another $n$ real
parameters. Finally, since there is no bare mass for the active
neutrinos, we have $y^*m_{\nu}y^{\dagger}=0_n$ which implies $\VC
m_{\nu}\VCT=0_n$ \cite{pilaftsis}. The constraint is symmetric,
and thus imposes $2n$ real conditions from the diagonal and
$n(n-1)$ from the off-diagonal. Thus, the number of real physical
parameters in $\VC$ is $2n(n_0-1)$. However, this counting is not
as advantageous as the above one. It does not distinguish between
mixing angles and CP phases, or tell how many of them enter in
$\VN$. More importantly, it does not take into account the further
reduction of physical parameters due to the appearance of massless
neutrinos when $n>n_0$.

The counting of physical parameters is summarized in Table 1 where
the result of Ref. \cite{sv80a} is also shown for comparison. The
difference arises from the fact that the zero bare mass for active
neutrinos has been completely exploited here to remove all
unphysical parameters while it was only partially applied in Ref.
\cite{sv80a} to delete unitary transformations within the massless
neutrinos for the case $n>n_0$.
\begin{table}
\begin{center}
\caption{\small The numbers of independent physical parameters are
shown for the mixing matrices $\VC$ and $\VN$. Note that all
parameters in $\VN$ are already included in $\VC$. Also shown are
the results of Ref. \cite{sv80a} for $\VC$.}
\begin{tabular}{|ll|l|l|}
\hline
&&$n_0\ge n\ge 1$&$n\ge n_0\ge 1$\\
\hline
$\VC$:&angles&$n(n_0-1)$&$n_0(n-1)$\\
      &phases&$n(n_0-1)$&$n(n_0-1)$\\
      &total &$2n(n_0-1)$&$2n_0n-(n+n_0)$\\
\hline
$\VN$:
      &angles&$n_0n-\frac{1}{2}n(n+1)$&$\frac{1}{2}n_0(n_0-1)$\\
      &phases&$n_0n-\frac{1}{2}n(n+1)$&$\frac{1}{2}n_0(n_0-1)$\\
      &total &$2n_0n-n(n+1)$&$n_0(n_0-1)$\\
\hline\hline
$\VC$:
      &angles&$n_0n+\frac{1}{2}n(n-1)$&
      $2n_0n-\frac{1}{2}n_0(n_0+1)$\\
\cite{sv80a}
      &phases&$n_0n+\frac{1}{2}n(n-1)$&
$2n_0n-n-\frac{1}{2}n_0(n_0-1)$\\
      &total &$2n_0n+n(n-1)$&$4n_0n-n-n_0^2$\\
\hline
\end{tabular}
\end{center}
\end{table}

\section{Approximate results for two sterile neutrinos}

The number of active neutrinos has been constrained by experiments
to be $n=3$. To accommodate neutrino mass and mixing, one would
introduce sterile neutrinos as few as possible. For $n_0=1$, we
have two massless neutrinos and two massive ones; and there are
two mixing angles in $\VC$. This has been shown in Refs.
\cite{liao,aguila} to be even impossible to explain the solar and
atmospheric experiments that call for two mass squared differences
($\Delta m^2_{ji}=m^2_j-m^2_i$) and two mixing angles, let alone
the LSND results hinting at a third $\Delta m^2_{ji}$. The next
simplest is the case with two sterile neutrinos which is studied
in this and the next section. Now we have one massless neutrino
($\nu_1$) and four massive ones ($\nu_{2,\cdots,5}$). Separating
out the massless mode from the matrix $D_{\triangle}$, we have
\begin{equation}
D_{\triangle}=\left(\begin{array}{cc} 0&d_2\\d_1&z
\end{array}\right),~~
M_{\rm diag}=\left(\begin{array}{cc}r_1&\\&r_2\end{array}\right)
\end{equation}
where generally $d_{1,2}>0,~r_{1,2}>0$ and $z$ is complex. These
parameters are traded for the four neutrino masses
$m_{2,\cdots,5}$ and one mixing angle plus one CP phase in $\VN$.
Both the angle and the phase enter in $\VC$ which includes
additional three angles and two phases coming from the matrices
$V_1,~V_2$. We thus should have enough degrees of freedom to
accommodate oscillation experiments that require at least three
$\Delta m^2_{ji}$ and four mixing angles.

Although it is possible to diagonalize algebraically the mass
matrix $m_n$, it is not instructive for our phenomenological
analysis. Experimentally, we need three well-separated $\Delta
m^2_{ji}$ to explain the positive oscillation results, $\Delta
m^2_{\odot}\ll\Delta m^2_{\rm ATM}\ll\Delta m^2_{\rm LSND}$. Since
the massless neutrino $\nu_1$ is purely active and the solar and
atmospheric experiments are more in favor of active neutrino
mixing than the involvement of sterile neutrinos, the other two
mainly active neutrinos $\nu_{2,3}$ have to lie close to $\nu_1$
while the two mainly sterile neutrinos $\nu_{4,5}$ must be well
above it; i.e., $m_{4,5}\gg m_{2,3}$. Furthermore, since the
sterile mass terms are not constrained by the low energy
symmetries of SM, it is natural that they may be linked to some
new physics at a higher scale. Thus, we shall assume the hierarchy
$r_{1,2}\gg d_{1,2},|z|$ in our phenomenological analysis. As a
special case, we shall also consider the possibility of $z=0$ in
which analytic results can be easily worked out for masses and
mixing matrices without assuming the hierarchy $r_{1,2}\gg
d_{1,2}$.

\subsection{See-saw case: $r_{1,2}\gg d_{1,2},|z|$}

Excluding the massless neutrino, the mass matrix to be
diagonalized via $Y^Tm_nY={\rm diag}(m_2,\cdots,m_5)$ is
\begin{equation}
m_n=\left(\begin{array}{cccc} &&&d_2\\
&&d_1&z\\
&d_1&r_1&\\
d_2&z&&r_2
\end{array}\right)
\end{equation}
The eigenvalues exact to the second order in the expansion of
$d_{1,2},z$ over $r_{1,2}$ are found to be
\begin{equation}
\begin{array}{rcl}
m_{2,3}&=&\displaystyle\frac{1}{1+|\rho_{\mp}|^2}
\left|\frac{d_1^2}{r_1}+\frac{(z+d_2\rho_{\mp}^*)^2}{r_2}\right|\\
m_4&=&\displaystyle r_1+\frac{d_1^2}{r_1}\\
m_5&=&\displaystyle r_2+\frac{d_2^2+|z|^2}{r_2}
\end{array}
\end{equation}
where $m_2<m_3$, $m_2m_3=\frac{d_1^2d_2^2}{r_1r_2}$ and
\begin{equation}
\begin{array}{rcl}
\rho_{\pm}&=&\frac{1}{2\xi}
\left[-\zeta\pm\sqrt{\zeta^2+4|\xi|^2}\right]\\
\xi&=&d_2\left[d_1^2z^*+\frac{r_1}{r_2}z(d_2^2+|z|^2)\right]\\
\zeta&=&d_1^2(z^2+z^{*2})+\frac{r_1}{r_2}(|z|^4-d_2^4)+\frac{r_2}{r_1}d_1^4
\end{array}
\end{equation}
When $z$ is real, there is an additional useful relation,
\begin{equation}
m_2+m_3=\frac{d_1^2}{r_1}+\frac{d_2^2+z^2}{r_2}
\label{eq_z}
\end{equation}
The first two rows of $Y$ give the submatrix $\bar{y}$ in $y$.
Including the massless degree of freedom corresponding to $y_0=1$
after removing unphysical parameters, we have
\begin{equation}
y^*=\left(\begin{array}{lllll}
1&0&0&0&0\\
0&v_-\rho_-&v_+\rho_+&0&\frac{d_2}{r_2}\\
0&v_-&v_+&\frac{d_1}{r_1}&\frac{z}{r_2}
\end{array}\right)
\end{equation}
where the second and third columns are exact to the first order
and the last two to the second order, and
\begin{equation}
v_{\mp}=\frac{ie^{i\alpha_{\mp}}}{\sqrt{1+|\rho_{\mp}|^2}}
\end{equation}
where $2\alpha_{\mp}$ are the phases of the complex numbers,
$\left(\frac{d_1^2}{r_1}+\frac{(z+d_2\rho_{\mp}^*)^2}{r_2}\right)$.

Parameterizing the $3\times 3$ unitary matrix
$V=(V_{\alpha\beta})$ with $\alpha,\beta=e,\mu,\tau$, we have
\begin{equation}
\VC=\left(\begin{array}{lllll} V_{ee}
&v_-(V_{e\mu}\rho_-+V_{e\tau})&
v_+(V_{e\mu}\rho_++V_{e\tau})&V_{e\tau}\frac{d_1}{r_1}&
(V_{e\mu}\frac{d_2}{r_2}+V_{e\tau}\frac{z}{r_2})\\
V_{\mu e} &v_-(V_{\mu\mu}\rho_-+V_{\mu\tau})&
v_+(V_{\mu\mu}\rho_++V_{\mu\tau})&V_{\mu\tau}\frac{d_1}{r_1}&
(V_{\mu\mu}\frac{d_2}{r_2}+V_{\mu\tau}\frac{z}{r_2})\\
V_{\tau e} &v_-(V_{\tau\mu}\rho_-+V_{\tau\tau})&
v_+(V_{\tau\mu}\rho_++V_{\tau\tau})&V_{\tau\tau}\frac{d_1}{r_1}&
(V_{\tau\mu}\frac{d_2}{r_2}+V_{\tau\tau}\frac{z}{r_2})
\end{array}\right)
\end{equation}
Since $y^{\dagger}y=1_3$ is exact at the first order in the
expansion, so is $\VC\VCd=1_3$.

\subsection{Case: $z=0$}
The neutrino mass matrix can be easily diagonalized in this case.
The masses are
\begin{equation}
m_{2,5}=\frac{1}{2}\left[\sqrt{r_2^2+4d_2^2}\mp r_2\right],~
m_{3,4}=\frac{1}{2}\left[\sqrt{r_1^2+4d_1^2}\mp r_1\right]
\end{equation}
There is no free angle or phase in the diagonalizing matrix $Y$
which is completely fixed by the masses, so that
\begin{equation}
y^*=\displaystyle\left(
\begin{array}{ccccc}1&&&&\\&is_2&0&0&c_2\\&0&is_1&c_1&0
\end{array}\right)
\end{equation}
where
\begin{equation}
\begin{array}{lcl}
c_1=\sqrt{\frac{m_3}{m_3+m_4}}&<&s_1=\sqrt{\frac{m_4}{m_3+m_4}}\\
c_2=\sqrt{\frac{m_2}{m_2+m_5}}&<&s_2=\sqrt{\frac{m_5}{m_2+m_5}}
\end{array}
\end{equation}
Note that the $i$ factors are introduced to make $m_{2,3}$ real
positive and do not by themselves signal CP violation
\cite{kayser}. Then
\begin{equation}
\VC=\left(
\begin{array}{ccccc}
V_{ee}&iV_{e\mu}s_2&iV_{e\tau}s_1&V_{e\tau}c_1&V_{e\mu}c_2\\
V_{\mu e}&iV_{\mu\mu}s_2&iV_{\mu\tau}s_1&V_{\mu\tau}c_1&V_{\mu\mu}c_2\\
V_{\tau
e}&iV_{\tau\mu}s_2&iV_{\tau\tau}s_1&V_{\tau\tau}c_1&V_{\tau\mu}c_2
\end{array}\right)
\end{equation}
We have $\VC\VCd=1_3$ exactly.

\section{Implications on neutrino oscillations}

We consider now the phenomenological implications of the previous
section. Before we move to neutrino oscillations, we discuss
briefly the results on the tritium decay \cite{mainz} and the
neutrinoless double $\beta$ decay \cite{nuless}. Both decays are
currently available experiments sensitive to the absolute neutrino
mass. The first one measures the following effective mass through
the distorted decay spectrum,
\begin{equation}
m^2_{\nu_e}=\sum_{j=1}^5m^2_j|\VC_{ej}|^2
=(Vy^*m^2_{\nu}y^TV^{\dagger})_{ee}
\end{equation}
Using $Y^Tm_nY=m_{\nu}$, $y^*m^2_{\nu}y^T$ is just the upper-left
$3\times 3$ submatrix of $m_nm^{\dagger}_n$, i.e.,
$D_{\triangle}D_{\triangle}^{\dagger}$. Thus the effective mass is
only sensitive to the Dirac mass terms,
\begin{equation}
m^2_{\nu_e}=d_2^2|V_{e\mu}|^2+(d_1^2+|z|^2)|V_{e\tau}|^2
+d_2(zV_{e\mu}^*V_{e\tau}+z^*V_{e\mu}V_{e\tau}^*)
\end{equation}
which is of order the product of a light neutrino mass and a heavy
one if $V_{ee}$ is not very close to unity.

The neutrinoless double $\beta$ decay violates the lepton number
conservation and can occur only when the neutrino is of Majorana
character. At the leading order in the expansion of neutrino mass
over the characteristic momentum transfer in the nuclear
transition, the decay is proportional to the effective mass
\begin{equation}
m_{ee}=|\sum_{j=1}^5m_j(V^{\rm C}_{ej})^2|
\end{equation}
Note that there is interference amongst different terms. In the
model considered here where the only extension is to add sterile
neutrinos to SM, the above mass vanishes. Actually we can show
that the effective mass for neutrinoless double like-sign charged
leptons ($\ell_{\alpha}^{\mp}\ell_{\beta}^{\mp}$) decays
\begin{equation}
m_{\alpha\beta}=|\sum_{j}m_jV^{\rm C}_{\alpha j}V^{\rm C}_{\beta
j}|
\end{equation}
vanishes generally for any number of generations and any number of
sterile neutrinos. Using $\VC=Vy^*$, we have
\begin{equation}
\sum_{j=1}^{n+n_0}m_jV^{\rm C}_{\alpha j}V^{\rm C}_{\beta j}
=\sum_{\gamma,\delta}V_{\alpha\gamma}V_{\beta\delta}
\sum_jm_jy^*_{\gamma j}y^*_{\delta j}
=\sum_{\gamma,\delta}V_{\alpha\gamma}V_{\beta\delta}
(y^*m_{\nu}y^{\dagger})_{\gamma\delta}
\end{equation}
where $(y^*m_{\nu}y^{\dagger})_{\gamma\delta}$ constitutes the
upper-left $n\times n$ submatrix of $Y^*m_{\nu}Y^{\dagger}=m_n$,
which vanishes however due to the absence of Majorana-type mass
terms for active neutrinos in the current model. While this does
not necessarily mean that such decays are forbidden, because they
can be induced at the next orders of the expansion and weak
interactions, it does imply they are strongly suppressed. It could
also be the case that one or more neutrinos are much heavier than
the momentum transfer so that the expansion does not apply to them
and the sum over $j$ is not complete. Even in this case, the
decays are still suppressed since the vanishing of the complete
sum and very large masses of those neutrinos imply very small
couplings between them and the charged leptons.

In the subsections to follow, we will consider the potential of
accommodating neutrino oscillation experiments in our simple
model. If the LSND signal does exist at all, it must point to a
much larger mass squared splitting than those in the solar and
atmospheric experiments. Since the lightest active neutrino is
massless in our model, it implies that the masses of the neutrinos
responsible for the LSND signal are much larger than those of
mainly active neutrinos involved in the solar and atmospheric
experiments. It is thus adequate to work in the see-saw limit
($r_{1,2}\gg d_{1,2},|z|$) to simplify our numerical analysis. We
will also consider a special case with $z=0$ in which analytic
results have been readily worked out in the last section.

\subsection{See-saw case: $d_1,d_2,|z|\ll r_1,r_2$}
\subsubsection{Including LSND and negative SBL results}

A difficulty to accommodate all oscillation results by adding
sterile neutrinos is the tension existing between the negative and
positive SBL experiments themselves. The statistical compatibility
of the two sides has been analyzed in Ref. \cite{sorel} amongst
others \cite{others}. It was found in Ref. \cite{sorel} that the
SBL results can be significantly better reconciled with two
sterile neutrinos than with a single one. Assuming the SBL
compatibility and CP conservation, they found the following
best-fit point with two sterile neutrinos, in our notation,
\begin{equation}
\begin{array}{rl}
(1)&m^2_4=0.92{\rm ~eV}^2,~m^2_5=22{\rm ~eV}^2;\\
&|\VC_{e4}|=0.121,~|\VC_{\mu 4}|=0.204,~|\VC_{e5}|=0.036,
~|\VC_{\mu 5}|=0.224
\end{array}
\label{eqn_fit1}
\end{equation}
Restricting the neutrino masses to sub-eV, the best-fit point
moves to
\begin{equation}
\begin{array}{rl}
(2)&m^2_4=0.46{\rm ~eV}^2,~m^2_5=0.89{\rm ~eV}^2;\\
&|\VC_{e4}|=0.090,~|\VC_{\mu 4}|=0.226,~|\VC_{e5}|=0.125,
~|\VC_{\mu 5}|=0.160
\end{array}
\label{eqn_fit2}
\end{equation}
The above results were obtained for a general model of three
active and two sterile neutrinos, in particular without parametric
constraints on the mixing matrix $\VC$. In our minimal model, the
independent physical parameters in $\VC$ are much less than the
general case as detailed in the last section; and in addition,
$\VC$ involves the neutrino mass ratios. It is thus interesting to
ask whether the above best-fit points can be accommodated with the
minimal number of physical parameters in the model.

With two sterile neutrinos, there are four mixing angles in
$\VC=Vy^*$, three of them from $V$ and one from $y^*$. Assuming CP
conservation, $z$ and $V$ become real so that no phase can appear.
Although the second and third columns of $y^*$ are pure imaginary,
their $i$ factors have no effect on oscillations. Since the above
best-fit points fall in the see-saw limit, it is good enough to
identify $m_4\approx r_1$, $m_5\approx r_2$ without loss of
generality. For the solar and atmospheric mass gaps that enter our
analysis through the matrix $y^*$, we use
\begin{equation}
\Delta m^2_{\odot}=8\times 10^{-5}{\rm ~eV}^2,~\Delta m^2_{\rm
ATM}=2.4\times 10^{-3}{\rm ~eV}^2 \label{eqn_fit3}
\end{equation}
A slight deviation from those numbers will not change our
qualitative conclusion. Since $m_3>m_2>m_1=0$, we must have
$m_3^2\approx\Delta m^2_{\rm ATM}$ while there are two possible
ways to arrange for the solar gap: either (a) $m_2^2\approx\Delta
m^2_{\odot}$ or (b) $\Delta m_{32}^2\approx\Delta m^2_{\odot}$.
Combining with the two best-fit points shown in eqns.
$(\ref{eqn_fit1},\ref{eqn_fit2})$, we have four schemes to
consider, named below as (1a), (1b), etc.

We find that schemes (1a,1b,2b) can be excluded without using any
constraints from the solar and atmospheric mixing angles. Take
scheme (1a) for instance. The best-fit values for the matrix
entries in eqn. $(\ref{eqn_fit1})$ translate into
\begin{equation}
\begin{array}{l}
|V_{e\tau}|\frac{d_1}{r_1}=0.121,~
|V_{\mu\tau}|\frac{d_1}{r_1}=0.204,\\
|V_{e\mu}\frac{d_2}{r_2}+V_{e\tau}\frac{z}{r_2}|=0.036,~
|V_{\mu\mu}\frac{d_2}{r_2}+V_{\mu\tau}\frac{z}{r_2}|=0.224
\end{array}
\label{eq_1a}
\end{equation}
The first two give
$\frac{d_1}{r_1}\ge\sqrt{0.121^2+0.204^2}=0.237$. Using eqn.
$(\ref{eq_z})$ we have the bound,
$\frac{z^2}{r_2}\le(\sqrt{m_2}-\sqrt{m_3})^2$, which implies then
$\frac{|z|}{r_2}\le\frac{|\sqrt{m_2}-\sqrt{m_3}|}{\sqrt{m_5}}=
5.85\times 10^{-2}$. Now we multiply the last one in eqn.
$(\ref{eq_1a})$ by $\frac{d_1}{r_1}$, use the second and
$\frac{d_1d_2}{r_1r_2}=\sqrt{\frac{m_2m_3}{m_4m_5}}=0.987\times
10^{-2}$ to arrive at the following result
\begin{equation}
\left|0.987\times 10^{-2}V_{\mu\mu} \pm
0.204\frac{|z|}{r_2}\right|=0.224\frac{d_1}{r_1}
\end{equation}
where the sign in the second term refers to $V_{\mu\tau}z$. Even
if the second term takes its largest magnitude of $1.20\times
10^{-2}$ and adds to the first one, it still requires
$|V_{\mu\mu}|\gg 1$ to reach the lowest value of the right-hand
side of $5.31\times 10^{-2}$. This scheme is thus excluded, and
similarly with the schemes (1b,2b).

For the scheme (2a), we have to appeal to a constraint from the
solar mixing angle. The best-fit matrix entries correspond to
\begin{equation}
\begin{array}{l}
|V_{e\tau}|\frac{d_1}{r_1}=0.090,~
|V_{\mu\tau}|\frac{d_1}{r_1}=0.226,\\
|V_{e\mu}\frac{d_2}{r_2}+V_{e\tau}\frac{z}{r_2}|=0.125,~
|V_{\mu\mu}\frac{d_2}{r_2}+V_{\mu\tau}\frac{z}{r_2}|=0.160
\end{array}
\label{eq_2a}
\end{equation}
The third one gives
\begin{equation}
\left|0.0262V_{e\mu} \pm
0.090\frac{|z|}{r_2}\right|=0.125\frac{d_1}{r_1}
\end{equation}
where $\frac{d_1}{r_1}\ge 0.243,~\frac{|z|}{r_2}\le 0.131$ and we
have used $\frac{d_1d_2}{r_1r_2}=2.62\times 10^{-2}$. Since the
best-fit solar angle $\theta_{\odot}$ is about $32^{\circ}$, using
$\VC_{ee}=V_{ee}\approx\sqrt{1-\sin^2\theta_{\odot}}$ we have
$|V_{e\mu}|\le s_{\odot}=\sin\theta_{\odot}\approx 0.53$. Then the
left-hand side in the above equation cannot exceed
$(1.39+1.18)\times 10^{-2}$, significantly deviating from the
lowest value of the right-hand side, $3.04\times 10^{-2}$. The
difficulty can be better viewed the other way around. Assuming
$V_{e\mu}^2+V_{e\tau}^2=s^2_{\odot}$ and using
\begin{equation}
\frac{d_2^2+z^2}{r_2^2}=
\frac{m_2+m_3}{m_5}-\left(\frac{d_1}{r_1}\right)^2\frac{m_4}{m_5}
\le 1.896\times 10^{-2}
\end{equation}
the LHS of the third equality in eqn. $(\ref{eq_2a})$ cannot
exceed the value,
$\sqrt{V^2_{e\mu}+V^2_{e\tau}}\sqrt{\frac{d_2^2+z^2}{r_2^2}}\le
0.138s_{\odot}$. Then, the equality requires that
$s_{\odot}\ge\frac{0.125}{0.138}=0.908$, or $\theta_{\odot}\ge
65^{\circ}$, which deviates far away from the best-fit solar
angle.

\subsubsection{Excluding negative SBL results}
Assuming the positive and negative SBL experimental results are
compatible, we have seen in the above that the best-fit points
obtained in Ref. \cite{sorel} cannot be accommodated in the
see-saw parameter region of the minimally extended model which has
much less physical parameters than a general nonrenormalizble
model. The deviations are so large that we suspect it may be
difficult to find a parameter region to encompass all SBL results
that is consistent with other oscillation experiments. Although
this issue should better be assessed by an extensive statistical
analysis, it goes beyond the scope of this work. Rather, we would
ask the question: since it is generally difficult to relax the
tension in SBL experiments, what will happen if we take seriously
the LSND signal but ignore the negative SBL results?

As the mass gap required by LSND is much higher than those by
other experiments, a simple but plausible way to answer the
question is to take the best-fit values for the latter and
consider whether there is any room for the LSND signal. Since only
LSND hints at a third mass gap besides the known solar and
atmospheric ones, it is natural to simplify our analysis further
by arranging either $r_1\ll r_2$ or $r_1\approx r_2$. In the first
case, the neutrino $\nu_5$ essentially decouples whose only effect
is to provide a small mass to $\nu_2$. The decoupling is verified
by the small entries in the 5th column of $\VC$. In the second
case, $\nu_{4,5}$ effectively form a Dirac neutrino when $m_4=m_5$
holds exactly. Even if they have a gap that happens to be a solar
or an atmospheric one, their contributions to other experiments
can still be ignored because of the smallness of the relevant
mixing entries.

We assume CP conservation so that $V$ and $z$ are real. For
definiteness, we assume $z>0$ so that
$\rho_+>0,\rho_-=-\rho_+^{-1}<0$. The conclusion does not change
for $z<0$. For oscillations, we can also ignore $i$ factors in the
2nd and 3rd columns of $\VC$. Then all matrix entries are
effectively real. The oscillation amplitude responsible for the
LSND signal appropriate for both cases of $r_1\ll r_2$ and
$r_1\approx r_2$ is
\begin{equation}
\begin{array}{rcl}
A_{\rm LSND}&=&4|\VCs_{\mu 4}\VC_{e4}+\VCs_{\mu 5}\VC_{e5}|^2\\
&=&
4\left[\left(\frac{d_1^2}{r_1^2}+\frac{z^2}{r_2^2}\right)V_{e\tau}V_{\mu\tau}
+\frac{d_2^2}{r_2^2}V_{e\mu}V_{\mu\mu}
+\frac{d_2z}{r_2^2}(V_{e\mu}V_{\mu\tau}+V_{e\tau}V_{\mu\mu})\right]^2
\end{array}
\end{equation}
which involves the upper-right $2\times 2$ submatrix in $V$. To
determine it, we apply the best-fit point
($\theta_{\odot}=32^{\circ},~\theta_{\rm
ATM}=45^{\circ},~\theta_{\rm Chooz}=0$) to the relevant $2\times
2$ submatrix in $\VC$:
\begin{equation}
\begin{array}{ll}
\VC_{e2}=\frac{V_{e\tau}\rho_+-V_{e\mu}}{\sqrt{1+\rho_+^2}}=s_{\odot},
&\VC_{e3}=\frac{V_{e\mu}\rho_++V_{e\tau}}{\sqrt{1+\rho_+^2}}=0,\\
\VC_{\mu 2}=\frac{V_{\mu\tau}\rho_+-V_{\mu\mu}}{\sqrt{1+\rho_+^2}}
=\frac{c_{\odot}}{\sqrt{2}}, &\VC_{\mu 3}
=\frac{V_{\mu\mu}\rho_++V_{\mu\tau}}{\sqrt{1+\rho_+^2}}
=\frac{1}{\sqrt{2}}
\end{array}
\end{equation}
which solves the $V$ entries in terms of $\theta_{\odot}$ and
$\rho_+$. The amplitude becomes
\begin{equation}
A_{\rm LSND}
=2s_{\odot}^2\left[\left(\frac{d_1^2}{r_1^2}+\frac{z^2}{r_2^2}\right)
\frac{\rho_+(1+c_{\odot}\rho_+)}{1+\rho_+^2}
+\frac{d_2^2}{r_2^2}\frac{c_{\odot}-\rho_+}{1+\rho_+^2}
+\frac{d_2z}{r_2^2}\frac{\rho_+^2-2c_{\odot}\rho_+-1}{1+\rho_+^2}\right]^2
\end{equation}

When $r_1\ll r_2$, we have approximately
\begin{equation}
m_2\approx\frac{d_2^2}{r_2},~m_3\approx\frac{d_1^2}{r_1},
~m_4\approx r_1,
~\rho_+\approx\frac{d_2z}{d_1^2}\frac{r_1}{r_2}\ll 1
\end{equation}
The amplitude is dominated by the first term
\begin{equation}
A_{\rm LSND}\approx 2s_{\odot}^2\frac{d_1^4}{r_1^4}\rho_+^2
\approx 2s_{\odot}^2\frac{d_2^2z^2}{r_1^2r_2^2} \approx
2s_{\odot}^2\frac{m_2}{r_2}\frac{z^2}{r_1^2}
%\approx 2s_{\odot}^2\rho_+^2
%\frac{\Delta m^2_{\rm ATM}}{\Delta m^2_{\rm LSND}}
\end{equation}
which is very small because $z\ll r_1(\sim 1{\rm ~eV~})\ll r_2$
and $m_2=\sqrt{\Delta m^2_{\odot}}<0.01{\rm ~eV~}$. For an order
of magnitude estimate, we note that $z$ should be about the same
order of $d_{1,2}$. Then, the order of $A_{\rm LSND}$ should be
roughly from $2s_{\odot}^2\frac{\Delta m^2_{\odot}}{\Delta
m^2_{\rm LSND}}\sim 4\times 10^{-5}$ to
$2s_{\odot}^2\sqrt{\frac{\Delta m^2_{\odot}\Delta m^2_{\rm
ATM}}{\Delta m^2_{\rm LSND}r_2^2}}\ll 2\times 10^{-4}$, which is
well below the required LSND level of $\sim 3\times 10^{-3}$. For
$m_4\approx r_1\approx r_2\approx\sqrt{\Delta m^2_{\rm LSND}}$, we
have the relations
\begin{equation}
\rho_+^{-1}-\rho_+\approx\frac{d_1^2-d_2^2+z^2}{d_2z},~
m_2m_4\approx\frac{(d_1^2+z^2)\rho_+^2+d_2^2-2d_2z\rho_+}{1+\rho_+^2}
\end{equation}
which, after some algebra, simplify considerably the amplitude to
\begin{equation}
A_{\rm LSND}\approx 2c_{\odot}^2s_{\odot}^2\frac{m^2_2}{m_4^2}
\approx 2c_{\odot}^2s_{\odot}^2\frac{\Delta m^2_{\odot}}{\Delta
m^2_{\rm LSND}}\sim 3\times 10^{-5}
\end{equation}
which is again far below the desired level.

In summary, even if we ignore the negative SBL results to avoid
the potential incompatibility in SBL experiments, we cannot
naturally accommodate simultaneously the results of solar,
atmospheric, reactor and accelerator experiments including LSND in
this model with see-saw parameters $d_1,d_2,|z|\ll r_1,r_2$.

\subsection{Case: $z=0$}

As we argued in the above, the see-saw hierarchy $r_{1,2}\gg
d_{1,2},|z|$ is natural both theoretically and phenomenologically.
The conclusion reached in the last subsection is thus quite
general. Nevertheless, as a good illustration we consider below a
special case where all relevant quantities can be calculated
without working perturbatively. Since
$\displaystyle{\prod_{j=2}^5}m^2_j=\det
(m_n^{\dagger}m_n)=(d_1d_2)^4$, neither of $d_{1,2}$ can vanish.
We therefore study the case of $z=0$ in a way parallel to the last
subsection.

\subsubsection{Including LSND and negative SBL results}

Although $m_3<m_4,~m_2<m_5$ but otherwise arbitrary, theoretical
and phenomenological considerations require that $r_{1,2}\gg
d_{1,2}$. Thus we generally have $m_{4,5}\gg m_{2,3}$. As the
problem is symmetric under the simultaneous interchanges
$\nu_2\leftrightarrow\nu_3$ and $\nu_4\leftrightarrow\nu_5$, we
can choose without loss of generality $m_5\ge m_4$. Since the
relative ordering of $m_{2,3}$ is still free, we have four schemes
to arrange for the solar and atmospheric mass gaps:
\begin{equation}
\begin{array}{clclrcl}
{\rm (S1)}&m_2^2&=&\Delta m^2_{\odot},&m_3^2&=&\Delta m^2_{\rm
ATM};\\
{\rm (S2)}&m^2_{32}&=&\Delta m^2_{\odot},&m_3^2&=&\Delta m^2_{\rm
ATM};\\
{\rm (S3)}&m_3^2&=&\Delta m^2_{\odot},&m_2^2&=&\Delta m^2_{\rm
ATM};\\
{\rm (S4)}&m^2_{23}&=&\Delta m^2_{\odot},&m_2^2&=&\Delta m^2_{\rm
ATM}
\end{array}
\label{eqn_schemes}
\end{equation}
The parameters $c_j,~s_j$ are collected in Table 2 that are
computed for the best-fit values shown in eqn. $(\ref{eqn_fit3})$
and the best-fit values $m^2_4,~m^2_5$ of Ref. \cite{sorel} shown
in eqs. $(\ref{eqn_fit1},\ref{eqn_fit2})$. A slight shift of the
best-fit values does not change much those parameters because of
the double square root dependence, and in particular our
qualitative conclusion below is rather stable.

It is clear from the table that it is far from possible to get
close to the large best-fit values for $|\VC_{\mu 4}|,~|\VC_{\mu
5}|$ in eqn. $(\ref{eqn_fit1})$. For the sub-eV fit, schemes
(S2,S4) seem to have a chance, so let us take a closer look. We
find $|V_{e\tau}|\approx 0.35,~|V_{e\mu}|\approx
0.56,~|V_{\mu\tau}|\approx 0.87,~|V_{\mu\mu}|\approx 0.73$. The
unitarity of $V$ is violated by
$|V_{\mu\tau}|^2+|V_{\mu\mu}|^2\approx 1.3>1$. More seriously,
large $|V_{\mu\tau}|$ and $|V_{\mu\mu}|$ imply a very small
$|V_{\mu e}|=|\VC_{\mu 1}|$ which cannot be tolerated by the data
of $\Delta m^2_{\rm ATM}$ oscillations; the circumstance with
$|V_{e\tau}|$ and $|V_{e\mu}|$ is conflicting: while their
intermediate size implies a too large $|V_{ee}|$ that is excluded
by Chooz, they are not large enough to explain $\Delta
m^2_{\odot}$ oscillation data. We can thus safely conclude that
the best-fit points of Ref. \cite{sorel} cannot be realized in the
current model.
\begin{table}
\begin{center}
\caption{The parameters $c_j,~s_j$ are computed for the best-fit
values $\Delta m^2_{\odot}=8\times 10^{-5}{\rm ~eV}^2$, $\Delta
m^2_{\rm ATM}=2.4\times 10^{-3}{\rm ~eV}^2$ and the best-fit
values $m^2_4,~m^2_5$ shown in eqs.
$(\ref{eqn_fit1},\ref{eqn_fit2})$, corresponding to the left and
right part of the table.}
\begin{tabular}{|l|l|l|l|l||l|l|l|l|}
\hline
schemes&$c_1$&$s_1$&$c_2$&$s_2$&$c_1$&$s_1$&$c_2$&$s_2$\\
\hline (S1)&$0.22$&$0.98$&$0.04$&$1.00$&$0.26$&$0.97$&$0.10$&$0.99$\\
\hline (S2,S4)&$0.22$&$0.98$&$0.10$&$0.99$&$0.26$&$0.97$&$0.22$&$0.98$\\
\hline (S3)&$0.10$&$0.99$&$0.10$&$0.99$&$0.10$&$0.99$&$0.22$&$0.98$\\
\hline
\end{tabular}
\end{center}
\end{table}

\subsubsection{Excluding negative SBL results}

Now we ignore the negative SBL experiments and consider the
consistency to accommodate all other experiments in the current
model. It is sufficient to restrict ourselves to those that can be
reasonably well described by vacuum oscillations, i.e., KamLAND,
K2K, Chooz, and LSND that cover all three mass gaps. We take as
our input the mass gaps shown in eqn. $(\ref{eqn_fit3})$ and
$\Delta m^2_{\rm LSND}\approx 1.2{\rm ~eV}^2$, and study the
possibility to obtain desired oscillation amplitudes for the
mentioned experiments. Again we consider the natural case
$m_{4,5}\gg m_{2,3}$ so that $\nu_{4,5}$ are mostly sterile and
$\nu_{2,3}$ mostly active, and simplify the matter further by
working with the `only three gaps', i.e., either
$m_4^2=m_5^2=\Delta m^2_{\rm LSND}$ or $m_5\gg m_4=\Delta m^2_{\rm
LSND}$. The masses $m_{2,3}$ are classified into the four schemes
shown in eqn. $(\ref{eqn_schemes})$.

For $m_4=m_5$, schemes (S1) and (S3) are equivalent because they
are related by a renumbering of indices; similarly with (S2) and
(S4). Thus we only need to consider the schemes (S1,S2) below. The
masses yield
\begin{equation}
\begin{array}{rl}
{\rm (S1)}:&c_1\approx 0.21,~s_1\approx 0.98,~c_2\approx
0.09,~s_2\approx 1.0\\
{\rm (S2)}:&c=c_1\approx c_2\approx 0.21,~s=s_1\approx s_2\approx
0.98
\end{array}
\end{equation}
The oscillation amplitudes that are multiplied to the oscillating
factors are, in scheme (S1):
\begin{equation}
\begin{array}{rlcl}
{\rm LSND}:&4|\VCs_{\mu 4}\VC_{e 4}+\VCs_{\mu 5}\VC_{e 5}|^2
&\approx&4|V_{e\tau}V^*_{\mu\tau}c_1^2+V_{e\mu}V^*_{\mu\mu}c_2^2|^2\\
{\rm Chooz}:&4|\VC_{e 3}|^2(|\VC_{e 1}|^2+|\VC_{e 2}|^2)
&\approx&4|V_{e\tau}|^2(1-|V_{e\tau}|^2)s_1^2\\
{\rm K2K}:&4|\VC_{\mu 3}|^2(|\VC_{\mu 1}|^2+|\VC_{\mu 2}|^2)
&\approx&4|V_{\mu\tau}|^2(1-|V_{\mu\tau}|^2)s_1^2\\
{\rm KamLAND}:&4|\VC_{e 1}|^2|\VC_{e 2}|^2&\approx&
4|V_{ee}|^2|V_{e\mu}|^2
\end{array}
\end{equation}
and in scheme (S2):
\begin{equation}
\begin{array}{rlcl}
{\rm LSND}:&4|\VCs_{\mu 4}\VC_{e 4}+\VCs_{\mu 5}\VC_{e 5}|^2
&\approx&4|V_{ee}V_{\mu e}|^2c^4\\
{\rm Chooz}:&4(|\VC_{e2}|^2+|\VC_{e3}|^2)|\VC_{e1}|^2
&\approx&4|V_{ee}|^2(1-|V_{ee}|^2)s^2\\
{\rm K2K}:&4(|\VC_{\mu 2}|^2+|\VC_{\mu 3}|^2)|\VC_{\mu 1}|^2
&\approx&4|V_{\mu e}|^2(1-|V_{\mu e}|^2)s^2\\
{\rm KamLAND}:&4|\VC_{e2}|^2|\VC_{e3}|^2&
\approx&4|V_{e\mu}V_{e\tau}|^2s^4
\end{array}
\end{equation}

Consider scheme (S1) first. Positive K2K and negative Chooz
results imply $|V_{\mu\tau}|^2\sim|V_{\tau\tau}|^2\sim\frac{1}{2}$
and tiny $|V_{e\tau}|^2$. The order $10^{-3}$ exclusion level at
Chooz restricts the $c_1^2$ term for LSND to be less than $\sim
10^{-6}$, thus subdominant to the $c_2^2$ term, which however
cannot exceed $c_2^4\sim 6.6\times 10^{-5}$, well below the
desired LSND signal. Scheme (S2) is not better. K2K and Chooz
imply $|V_{\mu e}|^2\sim\frac{1}{2}$ and tiny $|V_{ee}|^2$. But in
that case, the LSND result should be more negative than the Chooz
because the ratio of their amplitudes is, $\frac{\rm LSND}{\rm
Chooz}\approx\frac{c^4}{s^2}|V_{\mu e}|^2\sim 10^{-3}$, in sharp
conflict with the observations.

When $m_5\gg m_4\gg m_{2,3}$, we have $c_2\approx 0,~s_2\approx 1$
and $\nu_5$ decouples from the oscillation. Since
$m_2\approx\frac{d_2^2}{r_2}$, it is more appropriate to assume
$m^2_2=\Delta m^2_{\odot}$ than any other arrangements. Thus we
only have to consider the scheme (S1). This is just a special case
of the above discussions with the difference that the LSND signal
becomes more difficult to accommodate since it arises from the
first term that was ignored in the above.

\section{Conclusion}
If the LSND experiment is confirmed by MiniBooNE, the scenario of
three ordinary active neutrinos will be insufficient to explain
all neutrino oscillation data. A natural approach to the problem
is to add sterile neutrinos to allow for more independent mass
squared differences that could potentially accommodate three
well-separated mass gaps. We have investigated systematically the
simplest type of such models where the only extension to SM is the
addition of sterile neutrinos. In particular, we do not introduce
extra Higgs multiplets or higher dimensional effective operators
to induce masses for the active neutrinos, and the extended model
keeps renormalizable as SM. We found that for any numbers of
generations and sterile neutrinos its leptonic sector contains
much less physical parameters than previously expected, upon
exploiting completely the texture zero in the neutrino mass
matrix. The model thus becomes quite viable for phenomenological
analysis even if it contains two or three sterile neutrinos. We
demonstrated that the mixing matrix in the leptonic charged
current interactions has a factorized form and that the factor
containing the neutrino mass ratios makes the neutral current
interactions of neutrinos non-diagonal as well.

We have studied the phenomenological feasibility to accommodate
all oscillation results in the extended model with two sterile
neutrinos. We have restricted ourselves in this work to the
see-saw region of the parameter space which is most natural
considering the experimentally found well-separated mass gaps and
for which all relevant results can be worked out analytically. We
have used as our reference points the best-fit values of Ref.
\cite{sorel} based on the analysis of the complete SBL data.
Unfortunately, the answer is negative, and even more: even if we
take the LSND result seriously but ignore other null SBL results,
it is difficult to accommodate the best-fit values for the solar
(plus KamLAND), atmospheric (plus K2K), Chooz and LSND results in
the see-saw parameter region. A slight shift of the best-fit
values does not alter our qualitative conclusion. We attribute
this to a new feature of the minimally extended model exposed
here: the leptonic charged current mixing matrix is strongly
modified by mass ratios of neutrinos. Adding more free mixing
angles does not necessarily improve the simultaneous fitting of
masses and mixing angles. If the LSND signal is confirmed by
MiniBooNE, the see-saw region of the model with two sterile
neutrinos could largely be excluded.

Finally we discuss briefly how the current work could be extended.
Although the see-saw parameters are most natural from theoretical
point of view, it is highly desirable to make a complete scanning
of the parameter space. Since the mainly sterile neutrinos are
generally light or at least not very heavy in the model, fitting
data in regions other than the see-saw one may not cause too
serious fine-tuning of parameters. Suppose the LSND result will be
confirmed by MiniBooNE, a negative result of the scanning would
rule out the simplest extension of SM in its leptonic sector and
call for something really new to SM.

%\vspace{0.5cm}
%\noindent
%{\bf Acknowledgements}

%\newpage
%\baselineskip=20pt

\end{document}